\documentclass[aps,pre,reprint,floatfix,superscriptaddress]{revtex4-1}

\usepackage{graphicx}
\usepackage{amsmath,amssymb,amsfonts}
\usepackage{hyperref}
\usepackage{url}

\usepackage{breakurl}

\begin{document}

\title {Extracting correlations in earthquake time series using visibility graph analysis}

   \author{Sumanta Kundu}
   \email[ ]{sumanta@spin.ess.sci.osaka-u.ac.jp}
   \affiliation{
   \begin {tabular}{c}
    Department of Earth and Space Science, Osaka University, 560-0043 Osaka, Japan
   \end {tabular}}

   \author{Anca Opris}
   \email[ ]{anca\_opris@spin.ess.sci.osaka-u.ac.jp}
   \affiliation{
   \begin {tabular}{c}
    Department of Earth and Space Science, Osaka University, 560-0043 Osaka, Japan
   \end {tabular}}

   \author{Yohei Yukutake}
   \email[ ]{yukutake@onken.odawara.kanagawa.jp}
   \affiliation{
   \begin {tabular}{c}
      Hot Springs Research Institute of Kanagawa Prefecture, 586 Iriuda, Kanagawa Prefecture, Odawara 250-0031, Japan
   \end{tabular}}

   \author{Takahiro Hatano}
   \email[ ]{hatano@ess.sci.osaka-u.ac.jp}
   \affiliation{
   \begin {tabular}{c}
    Department of Earth and Space Science, Osaka University, 560-0043 Osaka, Japan
   \end {tabular}}

\begin{abstract}
   Recent observation studies have revealed that earthquakes are classified into several different categories.
   Each category might be characterized by the unique statistical feature in the time series,
   but the present understanding is still limited due to their nonlinear and nonstationary nature.
   Here we utilize complex network theory to shed new light on the statistical properties of earthquake time series.
   We investigate two kinds of time series, which are magnitude and inter-event time (IET), for three different categories
   of earthquakes: regular earthquakes, earthquake swarms, and tectonic tremors.
   Following the criterion of visibility graph, earthquake time series are mapped into a complex network by considering 
   each seismic event as a node and determining the links. As opposed to the current common belief, 
   it is found that the magnitude time series are not statistically equivalent to random time series.
   The IET series exhibit correlations similar to fractional Brownian motion for all the categories of earthquakes. 
   Furthermore, we show that the time series of three different categories of earthquakes can be distinguished 
   by the topology of the associated visibility graph. Analysis on the assortativity coefficient also reveals that 
   the swarms are more intermittent than the tremors. 
\end{abstract}

\maketitle

\section{Introduction}

\subsection{Network-theoretical time series analysis}

      Inspired by the exceptional success of the network theory in recent years~\citep{Albert2002,Newman2003,Newman2006,Boccaletti2006,Abe2004,Baiesi2004,Hope2015}, the analysis of time series from 
   the perspective of complex network has received considerable attention due to the standing requirement 
   of understanding the dynamical processes behind time series data~\citep{Zhang2006,YANG2008,
   Lacasa2008,Donner_2010,Gao_2016}. Often a real-world time series arises from nonlinear processes and 
   their precise identification is important for modeling purposes. Recently, a merging trend has been observed 
   coupling ideas both from the field of nonlinear time series analysis and complex network theory~\citep{ZOU2019}. 
   If a time series is mapped into a complex network, one may expect that such a network reflects some inherent 
   properties of the original time series. Thus, one can utilize the recent graph-theoretical tools to extract novel 
   properties hidden in the time series.
   
   Among several other methods~\citep{Donner_2010,Gao_2016}, the visibility graph~\citep{Lacasa2008} has 
   become popular due to its simplicity and wide range of applicability. This method has demonstrated its 
   potential in extracting several characteristic features of the time series such as the periodicity, fractality, 
   chaoticity, nonlinearity, and more~\citep{Lacasa2008,Lacasa2010,F_Donges2013}. A merit of the visibility 
   graph method is its ability to capture nontrivial correlations in nonstationary time series without introducing
   elaborate algorithms such as detrending. For instance, it has been shown that the visibility graph 
   corresponding to the time series generated from a fractional Brownian motion (fBm) is scale-free.
   Moreover, the exponent $\gamma$ for the degree distribution corresponds to the Hurst exponent ($H$)
   of the fBm as~\citep{Lacasa_2009}: 
   \begin{equation}
   \gamma = 3 - 2H.
   \label{eq:hurstexpo} 
   \end{equation}
   Since the fBm generates $f^{-\beta}$ power spectrum with $\beta = 1 + 2H$, the exponent 
   $\gamma$ of the visibility graph should correspond to $\beta$ as
   \begin{equation}
   \gamma = 4 - \beta.
   \label{eq:hurstexpo1} 
   \end{equation}
   The network-theoretical method enables us to estimate $H$ and $\beta$ more easily
   than other standard methods such as calculating power spectrum \citep{Lacasa_2009}.
   Therefore, it has been applied to extract the fBm-like nature of time series in several 
   contexts such as finance~\citep{YANG2009}, health science~\citep{Shao2010,Ahmadlou2010}, 
   image processing~\citep{Iacovacci2020}, and geophysics~\citep{Elsner2009,Donner2012}.
   
   In this paper, we study the nature of correlation in earthquake time series by means of visibility graph.
   In particular, we focus on the two important quantities: the magnitude and the inter-event time (IET) 
   between two consecutive earthquakes.
   
   \subsection{Characteristics of the seismic sequences: three categories of earthquakes}
   
   Thanks to the continuous progress in observation technologies, various kinds of earthquakes have been known to date.
   Aiming at the statistical characterization of earthquakes belonging to different categories,
   here we choose to analyze three well-established categories: regular earthquakes, earthquake swarms, and tectonic tremors.
   The fundamental difference among these three categories lies in their generation mechanisms
   and the time scale of energy release.
   
   A time series of regular earthquakes includes mainshock-aftershock sequences and the background activity. 
   While the latter is a Poissonian process, the former is generally clustered in space and time. Aftershocks are 
   triggered usually by the static stress change associated with the mainshock, as well as some other post-seismic 
   relaxation processes such as afterslip or fluid flow. Major fraction of the total energy is released almost 
   instantaneously at the time of the mainshock and slowly decreases in time. It is observed that the 
   magnitude-frequency distribution $P(M)$ obeys an exponential distribution, namely, the Gutenberg-Richter 
   (GR) law~\citep{GRlaw}: $P(M) \propto 10^{-bM}$, with $b$ taking a value around $1$ in the active fault zones
   \citep{Hatano2015}. On the other hand, the temporal decay of the frequency of aftershocks is described by 
   the Omori–Utsu law~\citep{Omori,Utsu1995}.    
   
      The same phenomenology is not observed for the other two categories of earthquakes. In contrast to 
   mainshock-aftershock sequence, a seismic swarm is defined as a cluster of earthquakes with similar 
   magnitudes, which usually occur in a volcanic or geothermal tectonic setting. The intrusion of fluids can 
   reduce the resistance of faults and redistribute the stress in such a manner that the energy is released 
   gradually and almost equally among the largest shocks~\citep{Hill1977}.
   The Omori-Utsu law does not generally hold for swarms.
   
   Tectonic tremors represent weak and repetitive seismic signals emitted from a plate boundary in 
   a subduction zone. To the current belief, fluids generated by slab dehydration may be a cause of 
   tremors~\citep{Obara2002}. Similar to swarm earthquakes, the tectonic tremor activity is characterized 
   by hypocentre migration but on a different spatial and temporal scale: tremors migrate up to several 
   hundreds kilometers, whereas swarms are more local. The statistical laws are largely unknown for tremors.
  
\subsection{Outline of the paper}  
    
      Based on the analysis of the visibility graph, we argue against the current popular belief that
      earthquake magnitude time series are indistinguishable from random time series.
      The same method is applied for the IET time series, showing fBm-like correlation clearly.
      We also show that the time series of three different types of earthquakes can be distinguished 
      in the topology of the associated visibility graph.
   
      The paper is organized as follows. We start by describing the visibility graph algorithm and the characteristics 
   of the three categories of earthquakes including the specifications of the studied seismogenic zones in Sec.\ 
   \ref{sec:method}. The existence of memory in the time series of magnitudes and IETs have been investigated in 
   Secs.\ \ref{sec:magnitude} and \ref{sec:IET}, respectively. We discuss the topology of the visibility graph for both 
   magnitude and inter-event time series in Sec.\ \ref{sec:netprop}. Finally, we summarize in Secs.\ \ref{sec:discussion} 
   and \ref{sec:summary}.  

\begin{figure}[t]
\begin{center}
\begin{tabular}{c}
\includegraphics[width=0.96\linewidth]{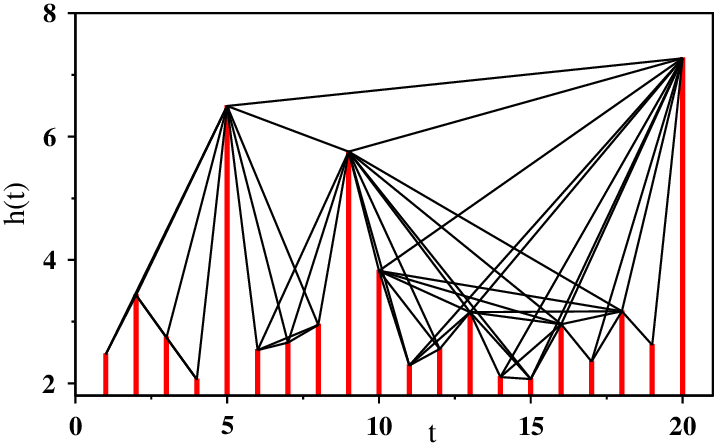}
\end{tabular}
\end{center}
\caption{Visibility graph representation of a synthetic time series with 20 data values drawn randomly from 
    an exponential distribution, where $t_i = i$ is the time $t$ corresponding to the $i$-th data. Each vertical 
    bar representing the height variable $h$ is considered as a node and if the top of one bar is 
    visible from the top of the another then a link is placed between the corresponding pair of nodes.
}
\label{fig:VGpic}
\end{figure}

\section{Methods}
\label{sec:method}

%
\begin{table*}[t]
\caption{The summary of the catalog data analyzed for investigating the correlations between the earthquake events.
}
\begin{tabular*}{\linewidth}{c@{\extracolsep{\fill}}lcccccccr}
\hline \hline \vspace{-0.27cm} \\
Earthquake type  & Region  & $\theta_{\rm min}$ &  $\phi_{\rm min}$  & $\theta_{\rm max}$  & $\phi_{\rm max}$  & Period  & $M_c$  & $N_t$   \\ \vspace{-0.27cm} \\ \hline \hline
Regular  & Tohoku               & 34.00     &  135.00  & 42.00    & 145.0    & 01/01/2000 -- 30/11/2019 & 2.0  & 147021 \\
         & Kumamoto               & 32.40     &  130.40  & 33.40    & 131.6  & 01/01/2000 -- 30/11/2019 & 1.0  & 44486 \\
         & Southern California & 30.00     & -124.00 & 39.00    & -111.0   & 01/01/1990 -- 08/12/2019 & 1.5  & 222491 \\ \hline 
Swarm    & Hakone              & 35.15     &  138.90  & 35.35    & 139.1  & 06/04/1995 -- 03/10/2015 & 0.1  & 16279   \\ 
                & Izu                     & 34.60    &  138.95   & 35.15    & 139.5 & 01/01/1995 -- 30/11/2019   & 0.0  & 38657 \\ \hline
Tremor   & Shikoku             & 33.66     &  131.61   & 34.28    & 134.5 & 01/04/2004 -- 01/09/2016 &        & 77701 \\
         & Cascadia                 & 37.50     &  -118.20  & 51.00    & -128.7 & 09/01/2005 -- 30/12/2014 &      & 30084 \\ \hline \hline
\end{tabular*}
\label{tab:catalog}
\end{table*}

\subsection{Construction of visibility graph from seismic catalog}

      Given the time sequence of the occurrence of seismic events, the visibility graph is constructed by considering each
   event as a node and linking the nodes based on mutual visibility of the corresponding data heights. The data recorded 
   at time $t_k$ is represented as the height $h_k$ of the $k$-th node. Specifically, any arbitrary pair of data values 
   $(t_i,h_i)$ and $(t_j,h_j)$ ($t_i<t_j$) are visible to each other if the straight line joining the two data points does 
   not intersect any intermediate data heights, as illustrated in Fig.\ \ref{fig:VGpic}.
   
   If there exists visibility, the slope $s_{ij}$ of the line between the nodes $i$ and $j$ must be the maximum of the slopes 
   $s_{ik}$ for all $i < k <j$. Therefore, a link is placed between two nodes $i$ and $j$ in the visibility graph if and 
   only if for all $t_i < t_k < t_j$ the following criterion is satisfied:
   \begin{equation}
   \label{criterion}
	   h_k < h_i + (h_j - h_i)\frac{t_k - t_i}{t_j - t_i}.
   \end{equation}
   Clearly, every node is visible at least from its left and right nearest neighbors and thus one obtains a completely 
   connected network.

      The ``divide \& conquer'' algorithm~\cite{Lan2015} has been used to efficiently transform a time series into its
   corresponding visibility graph. This algorithm takes advantage of the fact that the node with the maximum height
   divides the time series into two segments in the sense that the nodes situated at one side of the maximum are not visible
   from the another side. Therefore, it is not required to check the visibility between the two sides of each separated
   segments. In each step, the visibility of the node with the maximum height to the other nodes at its right and
   left sides is determined. Each new segment is then treated independently and the same procedure is repeated
   until every segment contains one single node. The CPU time taken by the algorithm scales with the size $N$ of a
   time series as $N\log N$.

\subsection{Description of the seismic catalog}

      In a seismic catalog, an event is described by the location of the hypocenter, the time of occurrence, and the magnitude 
   (M). We select several representative regions from Japan and California since these two areas are well-known
   for intense seismic activity and dense monitoring networks. The catalog data we analyze here are provided
   by the Japanese Meteorological Agency~\citep{jma}, the Hot Spring Research Institute~\citep{Yukutake2015}, 
   the Southern California Earthquake Center~\citep{scedc}, the World Tremor Database~\citep{tremor}, and Slow Earthquake 
   Database~\citep{Sloweq}, respectively.

      A selected region is described by the minimum and the maximum of the latitude ($\theta$) and longitude ($\phi$) 
   coordinates, i.e., the values of ($\theta_{\rm min},\phi_{\rm min}$) and ($\theta_{\rm max},\phi_{\rm max}$). We 
   consider only the crustal events within the depth of 50 km. For the regular and the swarm earthquakes, we also 
   indicate the magnitude of completeness $M_c$ i.e., the lowest magnitude above which the GR law holds. Above this 
   completeness magnitude, missing events in a catalog should be rare and therefore, effects of missing events should 
   be minimized. We determined these values using the Zmap software tool~\citep{Wiemer2001}. For tremors, we consider 
   all detected events recorded in the two previously mentioned database~\citep{Idehara2014,Mizuno2019}. The total 
   number of events in a catalog is denoted by $N_t$. The detailed specifications of these catalogs data are given 
   in Table~\ref{tab:catalog}.

\subsection{Remarks on regional specifics}

      For time series of regular earthquakes, we analyzed three active seismic regions located in different tectonic settings:
   subduction, compression, and active faulting. The region named Tohoku corresponds to an offshore 
   area of the Japan Trench subduction zone where the 2011 earthquake of moment magnitude M$_{\rm w}$9.0 and its aftershocks 
   were recorded. Time series before and after the M$_{\rm w}$9.0 event are referred here as Tohoku$_1$ and Tohoku$_2$, 
   respectively. The Southern California region is located in a complex compressional tectonic setting dominated by 
   the southern part of the San Andreas Fault system, but also includes earthquakes generated by the slow uplifting 
   of the Sierra Nevada Mountain range, as well as volcanic and geothermal related activity. The Kumamoto 
   region mostly includes the recent seismic activity generated by the 2016 M$_{\rm w}$7.0 Kumamoto earthquake 
   around the active Futagawa-Hinagu fault and the surrounding active volcanic region of Aso-Yufuin-Beppu. Thus, most earthquakes in the Kumamoto catalog are aftershocks. In the 
   Hakone volcanic region, significant swarm activity was detected since 2001~\citep{Honda2011}. Although many 
   different swarm episodes were recorded, they don't exhibit any specific temporal pattern. An increase in the 
   seismicity level was observed in 2015 due to a volcanic eruption~\citep{Yukutake2017}. The Izu volcanic region is characterized by magma-intrusion episodes  which generate frequent swarm activity~\citep{Hayashi2003}. Concerning the tremor 
   activity, we selected two areas where the largest number of detected events is available, such as Cascadia in 
   North America and Shikoku around the Nankai Trough in Japan.

\section{Analyses on magnitude time series}
\label{sec:magnitude}
\subsection{Stretched exponential nature of degree distribution}
%
\begin{figure}[t]
\begin{center}
\begin{tabular}{c}
\includegraphics[width=0.96\linewidth]{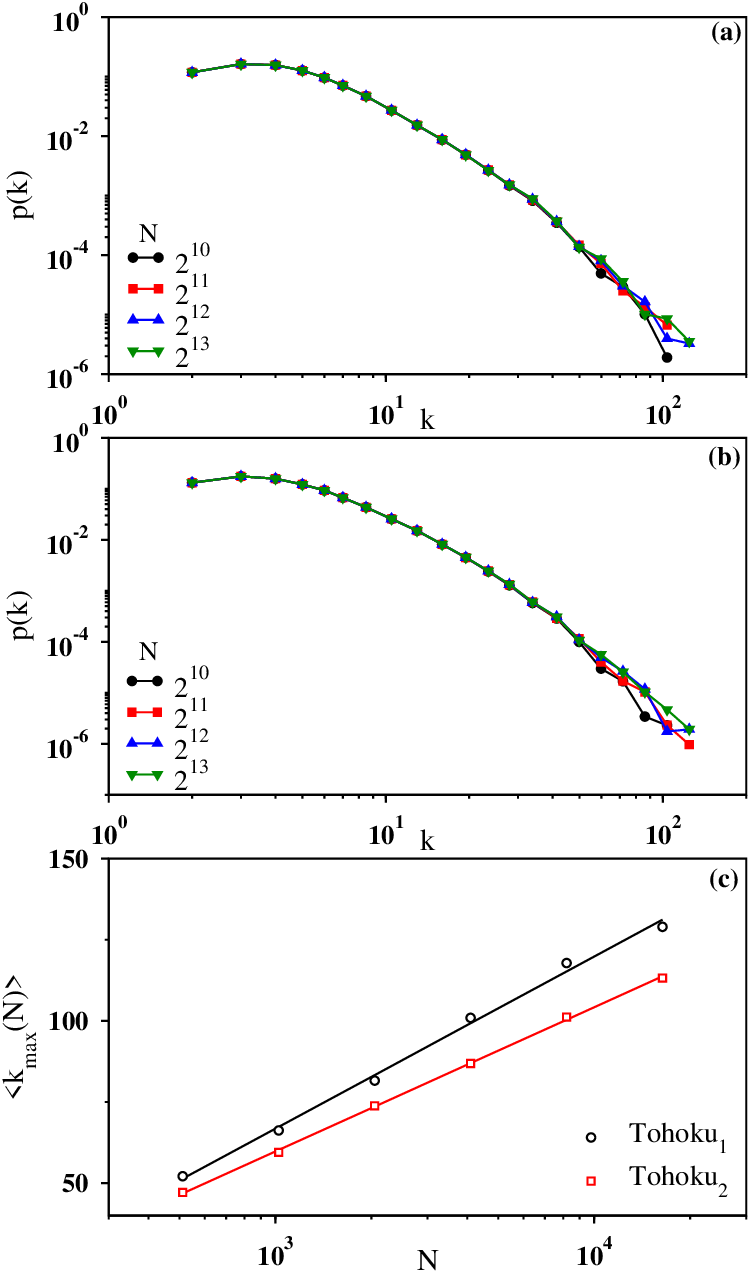} 
\end{tabular}
\end{center}
\caption{Log-log plot of the binned data for degree distribution $p(k)$ associated with the magnitude time series (a) Tohoku$_1$ and
(b) Tohoku$_2$ for network sizes $N = 2^{10}$ (black), $2^{11}$ (red), $2^{12}$ (blue), and $2^{13}$ (green). (c) The variation 
of the average maximum nodal degree $\langle k_{\rm max}(N) \rangle$ with $N$ on a lin-log scale for Tohoku$_1$ (black) and 
Tohoku$_2$ (red) using $N = 2^9$ to $2^{14}$. The fit (solid line) of the data points by a straight line indicating the logarithmic 
growth of $\langle k_{\rm max}(N) \rangle$.}
\label{fig:degdistscaling}
\end{figure}
%
\begin{figure}[tbh]
\begin{center}
\begin{tabular}{c}
\includegraphics[width=0.96\linewidth]{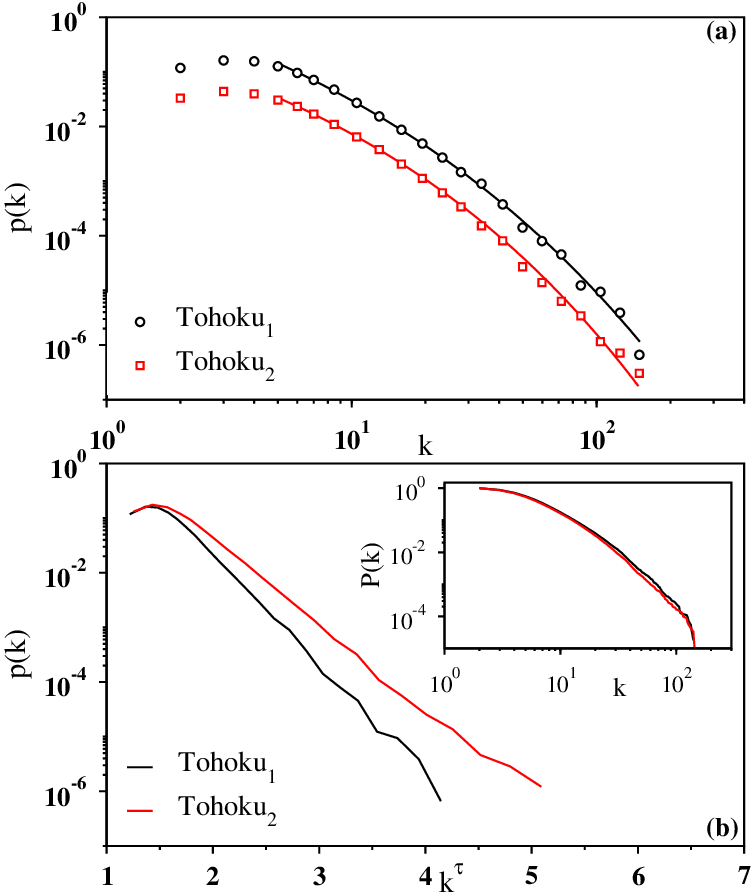}
\end{tabular}
\end{center}
\caption{(a) Log-log plot of the binned data (open circles) for degree distribution $p(k)$
of the whole magnitude time 
   series Tohoku$_1$ (black) and Tohoku$_2$ (red). The solid lines are the fit of the corresponding data using 
   Eq.\ (\ref{eq:stretchedexp}) whose parameters are $A$ = 195.0 and 57.68, $1/k_0$ = 210.0 and 51.03, and 
   $\tau$ = 0.284. and 0.325, respectively. The data for Tohoku$_2$ has been shifted vertically for visual clarity.
   (b) Plot of the same data against $k^{\tau}$, $k$ being the degree of the nodes, on a semilog scale exhibits a straight line in the intermediate regime. 
   Inset: log-log plot of the cumulative degree distribution $P(k)$ for the corresponding data sets.}
\label{fig:degdist}
\end{figure}
\begin{figure}[tbh]
\begin{center}
\begin{tabular}{c}
\includegraphics[width=0.96\linewidth]{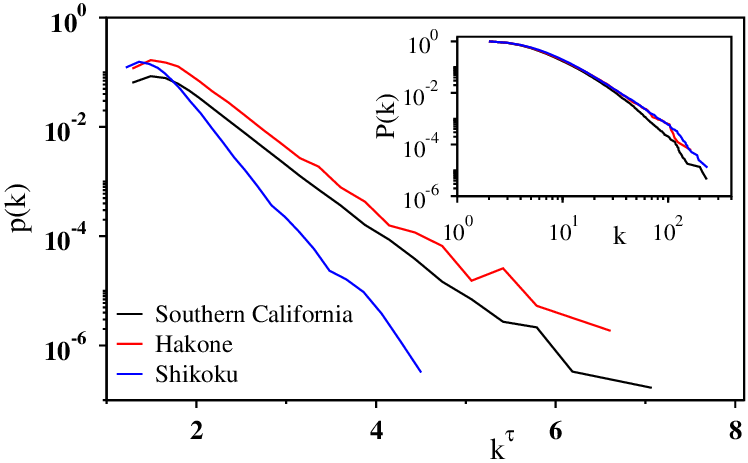}
\end{tabular}
\end{center}
\caption{Plot of the degree distribution $p(k)$ against $k^{\tau}$, $k$ being the degree of the nodes, on a semilog scale for the time series of Southern 
California (black), Hakone (red) and Shikoku (blue). The $\tau$ values are 0.364, 0.364, and 0.280, respectively. The 
plot indicates exponential decay of all the curves. For visual clarity, a linear shift is given to the black curve 
[$p(k) = p(k)/2$]. Inset: log-log plot of the cumulative degree distribution $P(k)$ for the corresponding data sets
displaying systematic curvatures of the curves.}
\label{fig:degdist_all}
\end{figure}

  To investigate whether the magnitude of earthquakes has any correlations, we study the degree distribution of 
  the visibility graph constructed from the magnitude time series.
  
  First we check if the degree distribution is power law.
  Typically, a power law distribution is characterized by a long tail that develops with the network 
  size $N$ in such a manner that the average maximum nodal degree $\langle k_{\rm max}(N) \rangle$ grows as 
  $\langle k_{\rm max}(N) \rangle \propto N^{\alpha}$. This signifies the existence of power-law degree 
  distribution for the infinitely large network, $N \to \infty$. In order to do this analysis, the original time 
  series is divided into several segments such that each segment contains exactly $N$ number of events. 
   
      We start with our results for regular earthquakes in the Tohoku region. Since the period of Tohoku$_2$ 
      is exceptionally active after the occurrence of the magnitude $9.0$ earthquake, we have analyzed the data 
      for Tohoku$_1$ and Tohoku$_2$ separately. In Figs.\ \ref{fig:degdistscaling}(a) and (b), the degree distribution 
      of the visibility graph is shown on a double logarithmic scale for four values of $N$ starting from $2^{10}$ to 
      $2^{13}$, at each step $N$ being increased by a factor of 2. For all the four values of $N$ in both the cases 
      (Tohoku$_1$ and Tohoku$_2$), the curves have certain amount of curvature and the tails of the degree 
      distributions do not elongate significantly as $N$ increases. To see this dependence more clearly, we have 
      plotted the average maximum nodal degree $\langle k_{\rm max}(N) \rangle$ against $N$ on a semilog scale 
      in Fig.\ \ref{fig:degdistscaling}(c). Clearly, this implies that $\langle k_{\rm max}(N) \rangle \sim \ln N$, 
      demonstrating that the degree distribution is not a power law: namely, the absence of fBm-like structure in the 
      magnitude time series.
      
      Specifically, the degree distribution appears to follow a stretched exponential function:
\begin{equation}
 p(k)=A e^{-\left(\frac{k}{k_0}\right)^\tau}
\label{eq:stretchedexp}   
\end{equation}      
   In Fig.\ \ref{fig:degdist}(a), we have plotted the degree distribution $p(k)$ of the visibility graph on a log-log 
   scale for the whole time series of Tohoku$_1$ and Tohoku$_2$ containing 55824 and 91197 events, 
   respectively. The logarithmically binned data for both the series fits quite well with the above functional form 
   in the range of $k$ between 6 to approximately 100. This is shown more explicitly in Fig.\ \ref{fig:degdist}(b),
   where $p(k)$ is replotted against $k^\tau$ on a semilog scale. The curves are straight in the intermediate 
   region, indicating that the distribution follows an exponentially decaying function of $k^\tau$. This behavior
   is also evident from the cumulative degree distribution shown in Fig.\ \ref{fig:degdist}(b) (inset).
   
   To confirm the ubiquity of the stretched exponential nature of degree distribution, we analyze the other 
   six earthquake catalogs. Figure \ref{fig:degdist_all} shows degree distributions presented similarly to 
   those in Fig.\ref{fig:degdist}(b) for Southern California (regular), Hakone (swarms), and Shikoku (tremors). 
   Apparently, these degree distributions are fitted with the stretched-exponential function irrespective of the 
   region or the earthquake type. The cumulative degree distributions are also shown in 
   Fig.\ \ref{fig:degdist_all} (inset).
   
   Furthermore, two important points should be remarked regarding the robustness of the above result.
   First, the stretched exponential nature does not significantly change even when the cutoff magnitude $M_c$
   is set to be lower or slightly higher than the completeness magnitude:
   Namely, the result is rather insensitive to some undetected smaller events. This may be 
   because the tail of the degree distribution is controlled by events of larger magnitude,
   which generally have higher visibility. Second, we confirm that the shape of degree distribution is unaltered
   even if the time series is with respect to the event index instead of the real occurrence time.
   Namely, the degree distribution remains stretched exponential even if 
   the event time $t_i$ is replaced by an integer $i$ in the visibility criterion, Eq. (\ref{criterion}).
%
\begin{figure}[tbh]
\begin{center}
\begin{tabular}{c}
\includegraphics[width=0.96\linewidth]{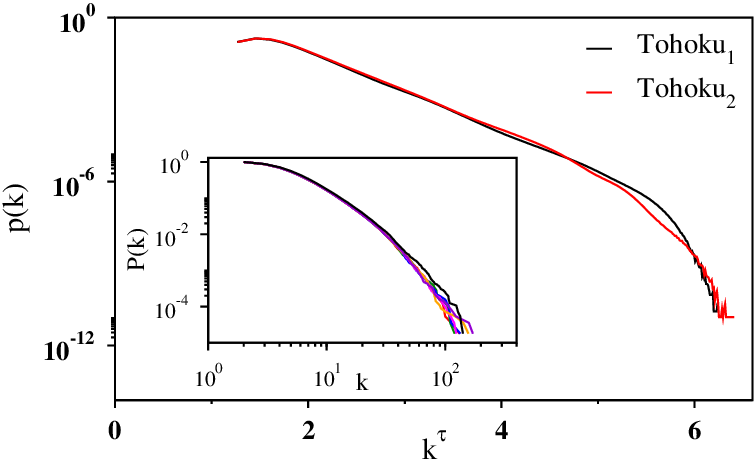}
\end{tabular}
\end{center}
\caption{Main panel: Semilog plot of the degree distribution, $p(k)$ vs $k^{\tau}$, $k$ being the degree of the nodes, for the shuffled 
   series corresponding to Tohoku$_1$ (black) and Tohoku$_2$ (red). The plot is based on $10^6$ 
   independent shuffled series. Inset: Log-log plot of the cumulative degree distribution $P(k)$ for 
   six individual shuffled series of Tohoku$_1$ (different colors are used to represent different 
   shuffled series) along with the original one (black).}
\label{fig:degdistSH}
\end{figure}
\begin{figure}[tbh]
\begin{center}
\begin{tabular}{c}
\includegraphics[width=0.96\linewidth]{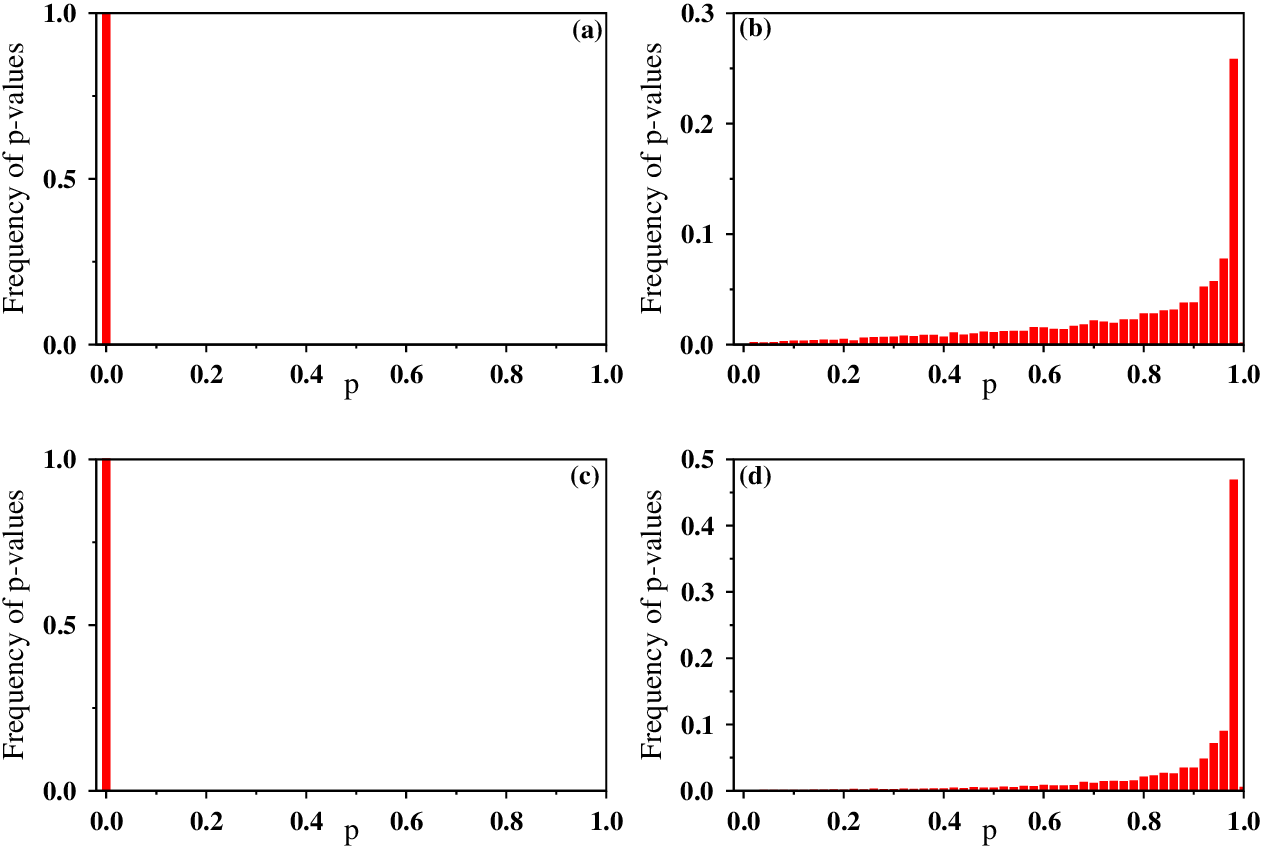}
\end{tabular}
\end{center}
\caption{Normalized frequency distribution of the $p$-values computed from the KS test statistic between
   two degree distributions: (Panels a and c) Comparison of degree distributions for the original time series and its shuffled ones.
   (Panels b and d) Comparison of degree distributions for two shuffled time series obtained from a given original time series.
   In each panel, the data is obtained from $10^4$ shuffled series.
   The upper and lower panel correspond to the time series of Tohoku$_1$ and Cascadia, respectively.}
\label{fig:KStest}
\end{figure}
 
\subsection{Degree distribution for shuffled data}

   Hereafter we refine the analysis and argue if there are any other correlations 
   in the magnitude time series. To this end, we first analyze the visibility graph produced from the 
   shuffled time series. Namely, by randomly choosing a pair of events, their respective magnitudes 
   are swapped. This process is repeated by $N_t$ times (the number of events in the catalog),
   leading to one shuffled sequence.
   This procedure preserves the probability density function of magnitude
   but destroys any potential correlations between them. 
   Then, for a shuffled sequence, the visibility graph is constructed and the degree distribution is calculated.
   This process is repeated for many times and the degree distributions are averaged over these shuffled sequences.
   The averaged degree distribution is shown in Fig.\ \ref{fig:degdistSH} (main panel).
   This is again fitted with the stretched exponential distribution. The same is true for the degree 
   distribution of each shuffled sequences, and the curves are not distinguishable from the original time series
   (inset of Fig.\ \ref{fig:degdistSH}). 
   This again validates the absence of fBm-like correlations in the original sequence.
   
\subsection{Kolmogorov-Smirnov test}

   More importantly, however, the above analysis does not mean that there are no correlations in earthquake magnitude,
   since the averaging process may mask some subtle short-range irregular correlations.
   To scrutinize the statistical difference in the visibility graph structure of the original and the shuffled
   time series, we perform the Kolmogorov-Smirnov (KS) test.
   Here the null hypothesis is that two empirical degree distributions originate from the same function for 
   the original time series and its shuffled sequence.
   We adopt the 0.05 significance level and reject this null hypothesis if the p-value is smaller than $0.05$.
   In this formulation, rejecting the null hypothesis means that the degree distributions are different for
   two visibility graphs produced from the original time series and its shuffled one.
   
  Specifically, the KS test statistic is computed as a distance between two empirical degree distributions
  produced from the original time series and its shuffled one. Then the p-value is calculated from the distance.
  This procedure is repeated for many shuffled sequences to yield the distribution of the p-value.
  They are shown in Figs.\ \ref{fig:KStest}(a) for Tohoku$_1$ (regular earthquakes) and (c) for Cascadia (tremors).
  Apparently, the null hypothesis is rejected for both the cases.
  Namely, the degree distributions are not the same for the original time series and the shuffled surrogates.
  We also find that the null hypothesis is rejected for all the other catalogs shown in Table I.
  This implies that the visibility structure in the original time series is somewhat altered if shuffled.
  In other words, the original time series can be discriminated among many other shuffled data.
  
  To support the above statement from another aspect, we again perform the KS test by comparing
  a specific shuffled time series with many other shuffled ones.
  The distribution functions for the p-value are shown in Figs.\ \ref{fig:KStest}(b) and (d).
  In this case, the null hypothesis is not rejected at the 0.05 significance level.
  Namely, shuffled time series are indistinguishable in terms of the degree distribution of their visibility graphs.
  This makes a quite contrast to the original time series, which is distinguishable from shuffled ones.
%
\begin{figure}[t]
\begin{center}
\begin{tabular}{c}
\includegraphics[width=0.96\linewidth]{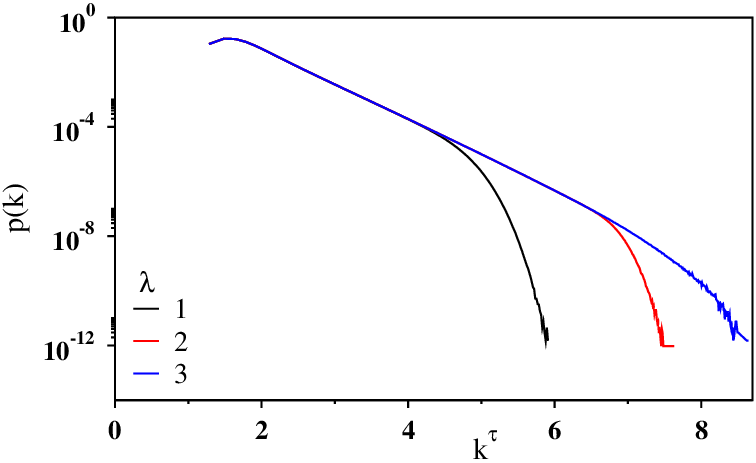}
\end{tabular}
\end{center}
\caption{Plot of the degree distribution $p(k)$ against $k^{\tau}$ with $\tau = 0.36$ for the visibility 
   graph associated with a random time series of $N = 2^{20}$ exponentially distributed data values 
   on a semilog scale for $\lambda$ = 1 (black), 2 (red), and 3 (blue).}
\label{fig:degdistEXP}
\end{figure}
\begin{figure}[tbh]
\begin{center}
\begin{tabular}{c}
\includegraphics[width=0.96\linewidth]{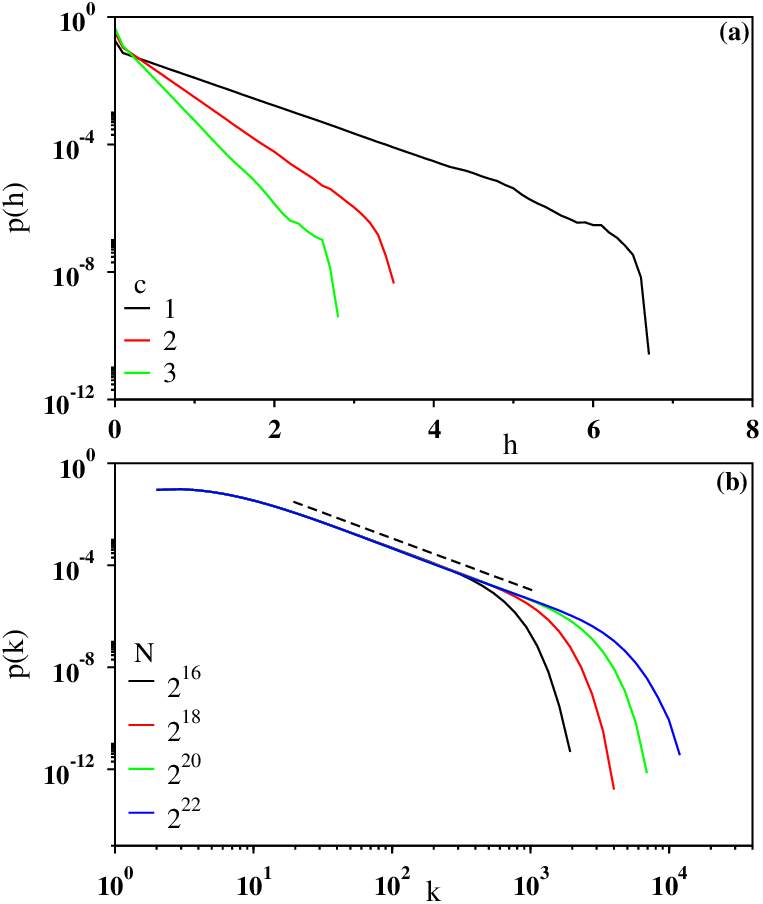} 
\end{tabular}
\end{center}
\caption{(a) Plot of the height distribution $p(h)$ of the time series generated from the Brownian motion 
   of a particle confined in a linear potential $U(x) = c |x|$ on a semilog scale for $c$ = 1 (black), 2 (red), 
   and 3 (green). The slopes of the curves are found to be 1.98(3), 3.95 (3), and 5.98(3), respectively. (b) 
   Log-log plot of the degree distribution $p(k)$ of the visibility graph corresponding to the 
   time series of $c = 1$ for $N$ = $2^{16}$ (black), $2^{18}$ (red), $2^{20}$ (green), and $2^{22}$ (blue). 
   The dotted line is the guide to the eye with slope 2.01. The results are based on the averages of at least 
   $10^3$ independent trajectories.
}
\label{fig:degdistBM}
\end{figure}
\subsection{Analyses on three other surrogates}   

   In addition to shuffled time series investigated above, we inspect three other surrogate data.
   The first and the second ones are the random time series, in which the height values $\lbrace h_i \rbrace$ are
   drawn randomly and independently from an exponential distribution $p(h) \sim e^{-\lambda h}$ between $[2, 9]$.
   For the first surrogate data, the time is set to be the event index: i.e., $t_i = i$ for the $i$-th event.
   Note that $\lambda$ is proportional to the $b$-value in the GR law as $\lambda = 2.303 b$.
   In Fig.\ \ref{fig:degdistEXP}, the degree distribution $p(k)$ are shown for several values of $\lambda$.
   Each curve is seen to follow the stretched exponential form. 
   Similar to the original earthquake data, $\langle k_{\rm max}(N) \rangle$ grows logarithmically
   with $N$ (not shown).
   This makes a contrast to the exponential degree distribution observed for the uniformly distributed heights~\citep{Lacasa2008}.

   The second surrogate data is the Poisson model, where events occur according to the Poisson process,
   and the height values are again drawn randomly from the GR law.
   We confirm that this surrogate data also produces the stretched exponential degree distribution.
   However, in the KS test that compares the surrogate data and the original magnitude series,
   the null hypothesis is rejected. Namely, they don't yield the same degree distribution.

   The third surrogate data we wish to inspect is a time series with a short memory.
   Here the time series is generated by simulating a Brownian particle in one dimension subjected to a linear potential:
   $U(x) = c |x|$. Starting from $x = 0$ at time $t = 0$, the position of the particle is updated in 
   steps of $dt = 10^{-6}$ according to the following Langevin equation:
   \begin{equation}
   x(t+dt) =
   \left\{
   \begin{aligned}
   & x(t) - cdt + \sqrt{dt} \xi  &\quad \text{for}~ x \geqslant 0, \\
   & x(t) + cdt + \sqrt{dt} \xi  &\quad \text{for}~ x < 0,
   \end{aligned}
   \right.
   \end{equation}  
   where $\xi$ is a Gaussian white noise with zero mean and unit variance.
   The height distribution for $x(t)$ follows the Boltzmann-Gibbs distribution at equilibrium.
   Since the potential is linear, the distribution function is exponential, as confirmed in Fig.\ \ref{fig:degdistBM}(a).
   We construct the visibility graph using the time series of $x(t)$ and compute the degree distribution.
   As shown in Fig.\ \ref{fig:degdistBM}(b) with four different system sizes $N$, the degree distribution
   is observed to follow a power law.
   Additionally, we confirm that $\langle k_{\rm max}(N) \rangle$ grows as a power-law with $N$: i.e., 
   $\langle k_{\rm max}(N) \rangle \sim N^{0.486(5)}$ (not shown).
   This signifies that a systematic single step memory in the time series leads to a scale-free network. 

   All the findings above lead us to conclude that the time series of earthquake magnitude are not statistically
   identical to uncorrelated time series, although no apparent systematic memories exist,
   either long-ranged (fBm-like) or short-ranged.
   
\begin{figure}[t]
\begin{center}
\begin{tabular}{c}
\includegraphics[width=0.96\linewidth]{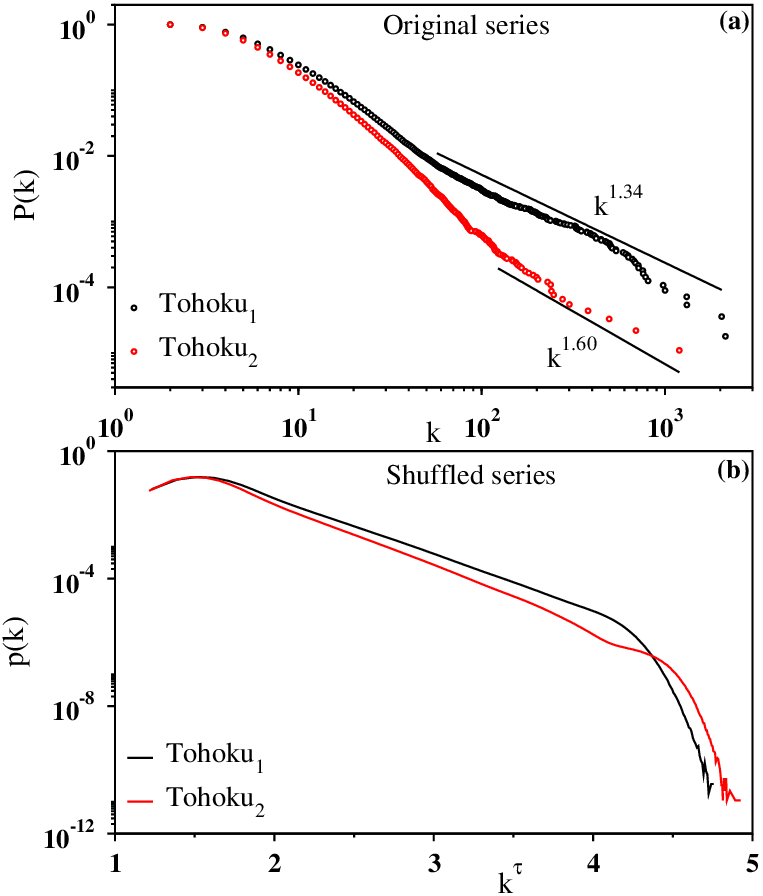} 
\end{tabular}
\end{center}
\caption{(a) Log-log plot of the cumulative degree distribution $P(k)$ for the IET series of Tohoku$_1$ (black) and Tohoku$_2$ (red).
The slope of the curve in the fitted region (solid line) has been estimated as 1.34(5) and 1.60(8), respectively.
(b) The degree distribution $p(k)$ shown as a function of $k^{\tau}$ with $k$ being the degree of the nodes for shuffled sequences of the corresponding data on a semi-log scale.
    Here the exponent $\tau$ is estimated as 0.30 and 0.28, respectively.}
\label{fig:IETdegdist-TH}
\end{figure}

\section{Correlation between the Inter-event times}
\label{sec:IET}
\subsection{Power-law nature of degree distribution for Tohoku data}

   To characterize the temporal correlations between seismic events and to understand whether they 
   are dependent on specific details of the seismic activity, we focus on studying the inter-event time 
   (IET) series of earthquakes. Here the IET series is obtained from an earthquake catalog
   by calculating time intervals between two consecutive events and labeling them with the event index $i$.
   Namely, the IET series is represented as $(i, h_i)$, where $h_i = t_{i+1} - t_i$, 
   and $t_i$ is the real occurrence time of the $i$-th event in a catalog. Here the threshold is set 
   as the completeness magnitude $M_c$ (listed in Table I).
   
   Fig.\ \ref{fig:IETdegdist-TH}(a) shows the cumulative degree distribution $P(k)$
   for the IET series of Tohoku$_1$ and Tohoku$_2$.
   This is the probability of finding a node with degree at least $k$ in the visibility graph.
   For both the cases, the degree distribution is found to be heavy-tailed distribution 
   and the tail more than one decade can be described by an approximate power law. We estimate the exponent:
   $\gamma = 2.34(5)$ for the Tohoku$_1$ and $\gamma=2.60(8)$ for the Tohoku$_2$. 
   The average maximum nodal degree also varies as a power law: 
   $\langle k_{\rm max}(N) \rangle \sim k^{\alpha}$, where $\alpha$ = 0.77(3) and 0.53(4) for the 
   Tohoku$_1$ and Tohoku$_2$, respectively (not shown here). This behavior supports the power law 
   nature of the degree distribution. Thus, the visibility graphs constructed from the IET series exhibit 
   typical signatures of a scale-free network, indicating the existence of fBm-like correlations in the 
   time series.

   To validate the presence of correlation in a contrasting manner, we analyze the shuffled 
   sequences of the IET data and find that the degree distribution $p(k)$ is fitted with the 
   stretched exponential function given in Eq.\ (\ref{eq:stretchedexp}). In Fig.\ \ref{fig:IETdegdist-TH}(b), 
   the degree distributions $p(k)$ are plotted with $k^{\tau}$ for the shuffled IET series 
   of Tohoku$_1$ and Tohoku$_2$. The straight line here confirms the stretched exponential form of 
   the degree distribution. In addition, we find that $\langle k_{\rm max}(N) \rangle \sim \ln N$
   (not shown). Evidently, the shuffled data produces the properties of a random time series and 
   therefore provides evidence on the existence of correlation in the original time series. 
 
\subsection{Power-law nature of degree distribution: other regions}
   
\begin{figure}[tbh]
\begin{center}
\begin{tabular}{c}
\includegraphics[width=0.96\linewidth]{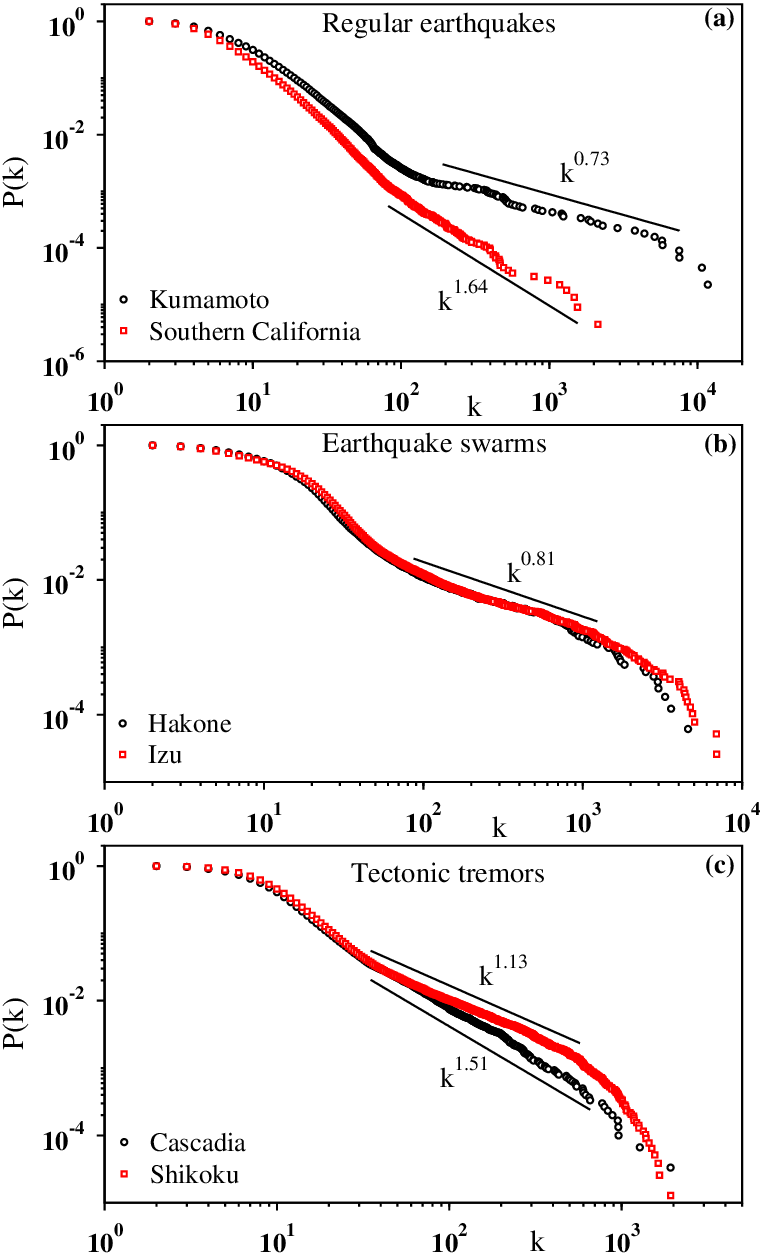}
\end{tabular}
\end{center}
\caption{Log-log plot of the cumulative degree distribution $P(k)$ for the IET series of different types of 
      earthquakes: (a) regular earthquakes in Kumamoto (black) and Southern California (red); (b) swarms 
      in Hakone (black) and Izu (red); (c) tremors in Cascadia (black) and Shikoku (red). The slopes in the 
      fitted region (solid line) are 0.73(8), 1.64(5), 0.81(9), 0.79(9), 1.51(5), and 1.13(5), respectively.}
\label{fig:IETdegdist}
\end{figure}

      The same analyses are carried out for regular earthquakes in different regions, as well as for 
   swarms and tremors. The results are shown in Fig.\ \ref{fig:IETdegdist}. In Figs.\ \ref{fig:IETdegdist}(a),
   (b) and (c), the cumulative degree distribution is plotted for regular earthquakes, swarms, and tremors. 
   For every case, a heavy-tailed distribution has been observed. While for regular earthquakes and 
   tremors a power law regime extending more than one decade is quite apparent, the data for swarms 
   shows more complex behavior. However, an approximate power law variation can fit the data in the 
   intermediate region. For each case, the data points in the most linear regime (estimated by eyes) 
   starting from a moderate value of $k$ to a value at the tail part upto which they do not fall-off 
   due to the limitations by finite size are fitted to the best straight line. From the slope of the 
   straight line, we estimate the power-law exponent $\gamma$ as 1.73(8), 2.64(5), 1.81(9), 
   1.79(9), 2.51 (5) and 2.13(5) for Kumamoto (regular), Southern California (regular), Hakone (swarm), 
   Izu (swarm), Cascadia (tremor), and Shikoku (tremor), respectively. In addition, the power law 
   dependence of the average largest degree $\langle k_{\rm max}(N) \rangle$ with $N$ has been 
   observed for every set of data (not shown), supporting the power law nature of the degree distribution.

%
%
   
   The tail part of the degree distribution is characterized by the exponent $\gamma$,
   which seems to depend on the seismic activity of the specific region:
   i) Earthquake swarms (Izu and Hakone) have a common value, $\gamma\simeq1.8$.
   ii) Regular earthquakes may also have a common value, $\gamma\simeq2.6$ (Tohoku$_2$ and Southern California), while it is somewhat smaller ($2.3$) before the Tohoku M$_{\rm w}$9.0 earthquake (Tohoku$_1$).
   iii) Kumamoto is exceptional with $\gamma\simeq 1.7$. This value is rather close to swarms, although the data mainly consist of aftershocks of 2016 Kumamoto earthquake.
   There may be two reasons for this discrepancy. First, the data is not a usual mainshock-aftershocks sequence, but rather a foreshocks-mainshock-aftershocks sequence. Alternatively, we may interpret it as two mainshocks (M$_{\rm w}$6.2 and 7.0) that occurred within only thirty hours. In any case, it is rather anomalous seismic activity. The second potential reason is an active volcano (Mt.\ Aso) located in the proximity of the main fault. The M$_{\rm w}$7.0 mainshock triggered many earthquakes in the volcanic area, including an M$_{\rm w}$5.9 event and its own aftershocks. Thus, the overall seismic activity is influenced by the nearby volcanic field and this may explain the resemblance to swarms.
   
   If we suppose the relation between the fractional Brownian motion and the power-law degree distribution,
   i.e., Eq.\ (\ref{eq:hurstexpo1}), the exponent for the power spectrum $\beta$ can be determined.
   For example, swarms have $\beta \simeq 2.2$ and $H\simeq0.6$.
   They are close to those for standard Brownian motion ($\beta=2$ and $H=0.5$) but yet slightly larger,
   corresponding to superdiffusion. Regular earthquakes ($\gamma\simeq2.6$) have $\beta=1.4$ and $H=0.2$,
   corresponding to subdiffusion. Extraction of these exponents from actual seismic data is difficult
   using other standard methods such as autocorrelation functions due to the strong nonstationary nature of the seismic record.
   In this sense, these exponents might not be considered as that for fBM itself, but should represent some counterpart in seismic activities.
   
   Shuffled time series again yield visibility graphs with their degree distributions of stretched-exponential form,
   resembling the properties of a random time series.
   Therefore, we confirm that the original IET series possess fBm-like correlations irrespective of the earthquake types:
   regular, swarms, and tremors.
   This result does not contradict the previous studies on regular earthquakes obtained using some different methods~\citep{Corral2006,Fan2019}.
   Here we have confirmed the correlation using complex network based approach, and more importantly, found correlations 
   in tremors and swarms.
   
   Lastly, we wish to add a remark on catalogs on tremors.
   Since a single event is not as distinct as regular earthquakes, there may be some errors in the IET of tremors.
  To check the effect of such errors in IET, we add a certain amount of noise to the IET data of tremors
  and construct the visibility graph from these noisy data.
  We find that the degree distribution is indeed robust to the noise: it retains the power-law nature against the small noise in IET.
   
%
\begin{figure}[tbh]
\begin{center}
\begin{tabular}{c}
\includegraphics[width=0.96\linewidth]{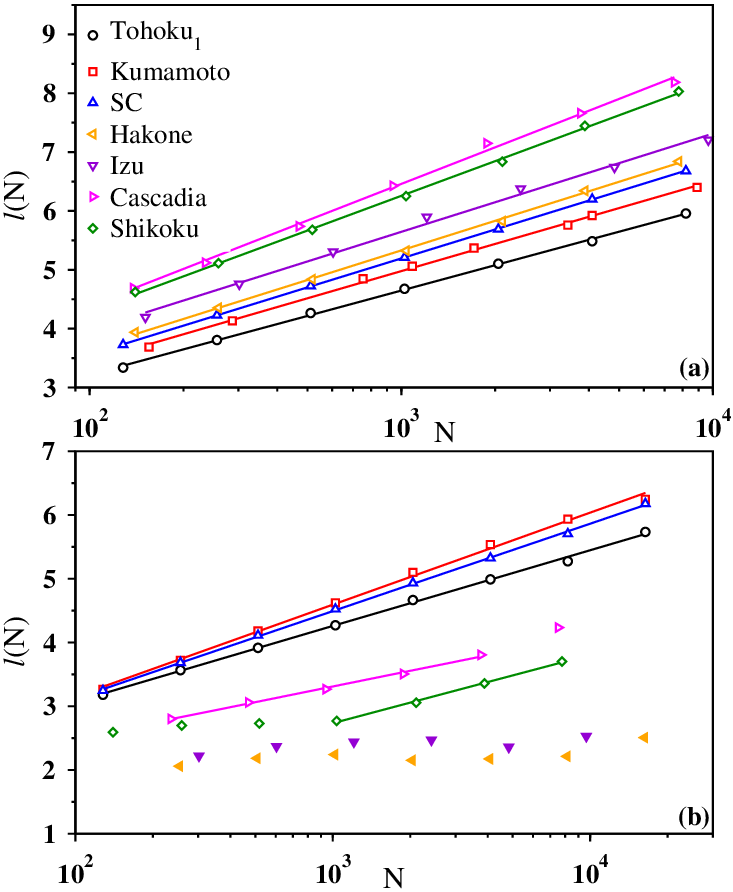}
\end{tabular}
\end{center}
\caption{The plots exhibit small-world behavior of the visibility graph for (a) magnitude and (b) IET
time series of Tohoku$_1$ (black), Kumamoto (red), Southern California (blue), Hakone (solid orange), Izu 
(solid violet), Cascadia (magenta) and Shikoku (green). For visual clarity data of $l(N)$ have been shifted 
vertically. Multiplicative factors in the upper panel are 1, 1.05, 1.10, 1.15, 1.20, 1.25, and 1.30, respectively.  
In the lower panel (the labels are same as in (a)) data for swarms have been shifted as $y=y/1.5$.
}
\label{fig:spath-TH}
\end{figure}

\section{Detailed structure of visibility graph}
\label{sec:netprop}

   The detailed characterization of the topology of the network has served to identify several non-trivial
   features exhibited by diverse types of real-world systems including the basic principles that played role 
   in the network formation~\citep{Newman2006,Albert2002,Boccaletti2006}. In order to extract more 
   properties hidden in the seismic records, the following graph-theoretical quantities have been analyzed 
   after obtaining the visibility graph using ``divide \& conquer'' algorithm.
   
      Since our visibility graph is connected and undirected, there always exists at least 
   one path between any arbitrary pair of nodes $i$ and $j$ through the links of intermediate nodes. The path with 
   the minimal links traversed is called the shortest path length $d_{ij}$, and the average shortest path length is 
   defined as,
\begin{equation}
l = \frac{1}{N(N-1)} \sum_{\substack{i, j \\ i \neq j}} d_{ij}.
\end{equation}
   In Figs.\ \ref{fig:spath-TH}(a) and (b), we show the variation of $l(N)$ with $N$ on a semilog scale for both the 
   magnitude and IET series, respectively. The best fit of the data by a straight line indicates its logarithmic scaling 
   and hence, the network is small-world. Although the data for IET series of Shikoku has some curvature, the linear 
   behavior is quite apparent for large values of $N$. For IET series of swarms, $l(N)$ grows more slower than $\ln N$.
   
\begin{figure}[tbh]
\begin{center}
\begin{tabular}{c}
\includegraphics[width=0.96\linewidth]{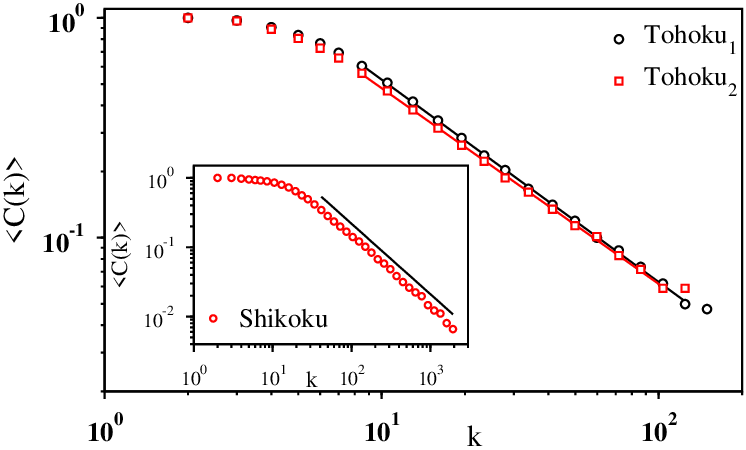}
\end{tabular}
\end{center}
\caption{The hierarchical nature of the visibility graph. The clustering coefficient $\langle 
C(k) \rangle$ has been plotted with $k$ on a log-log scale for the Tohoku$_1$ (black) and 
Tohoku$_2$ (red) magnitude time series (main panel), and for the IET series of Shikoku 
(inset). The slopes of the fitted lines have been measured as 0.92(3), 0.89(3), and 1.01(2), 
respectively. 
}
\label{fig:Hier}
\end{figure}
   
      Another important quantity associated with the network is the clustering coefficient which measures the three 
   point correlation among the neighbors. Specifically, the clustering coefficient $C_i$ of node $i$ measures the 
   probability that the two neighbors of $i$ are connected. If there exists $E_i$ links among the $k_i$ neighbors of 
   node $i$ then, $C_i = 2E_i / k_i(k_i - 1)$. In the case of $k_i < 2$, $C_i = 0$. The global clustering coefficient is 
   expressed as,
   \begin{equation}
   C = \langle C_i \rangle = \frac{1}{N} \sum_{i = 1}^{N} \frac{2E_i}{k_i(k_i - 1)}.
   \end{equation}
   By varying $N$ from $2^9$ to $2^{16}$ we have observed that $C$ is almost independent of $N$ (values differ
   only at 4-th decimal place) for both magnitude and IET time series of different types of earthquakes. Further, the 
   clustering coefficient $\langle C(k) \rangle$ for the nodes with degree $k$ has been found to decay as $\langle 
   C(k) \rangle \sim k^{-\nu}$ with $\nu \approx 1$, as shown in Fig.\ \ref{fig:Hier}. This is the universal feature of a 
   hierarchical network observed in many real-world networks~\citep{Ravasz2003}. The clustering coefficient $C$ 
   assumes its highest value for the IET series of tremors.

      We have also calculated the Pearson correlation coefficient $r$ to investigate whether a high degree node 
   tends to be linked with a high degree node (assortative mixing, $r > 0$) or a low degree node (disassortative 
   mixing, $r < 0$). We have calculated $r$ using the following formula~
   \citep{Newman2002},
\begin{equation}
r=\frac{L^{-1}\sum_ik_{1i}k_{2i}-[L^{-1}\sum_i\frac{1}{2}(k_{1i}+k_{2i})]^2}{L^{-1}\sum_i\frac{1}{2}({k_{1i}}^2+{k_{2i}}^2)-[L^{-1}\sum_i\frac{1}{2}(k_{1i}+k_{2i})]^2},
\end{equation}
   where, $k_{1i}$ and $k_{2i}$ are the degrees of nodes at the ends of link $i$ with $i = 1, 2, \cdots, L$. We found 
   that for all earthquake types, the magnitude series shows assortative nature (last column of Table \ref{tab:mag}). 
   In contrast, in case of IET series we obtain a value of $r \approx 0$ for the regular earthquakes and for swarms 
   and tremors $r < 0$ (last column of Table \ref{tab:IET}). Moreover, the graph associated with the IET series of 
   swarms has been found to be more disassortative than that of tremors. This means that for swarms the high 
   degree nodes show more preference towards linking with the low degree nodes. This indicates that the smaller 
   heights are abundant in both the time series, however, there are a few very large heights (i.e., long
   quiescence periods) in the swarms series which are even larger than the largest height in the tremor series. 
   Therefore, swarms are more intermittent than tremors. 

      For a detailed comparison of the characteristic differences among the three different types of earthquakes, the 
   above quantities have been calculated for a fixed value of $N = 2^{12}$ and the obtained values are listed in
   Table \ref{tab:mag} and Table \ref{tab:IET} for the magnitude and the IET series, respectively. Clearly, they can 
   be distinguished by the values of different graph-theoretical quantities obtained from their individual IET series.
      
\begin{table*}
	\caption{Average values of the maximum degree $k_{\rm max}$, average degree $\langle k \rangle$, clustering
	coefficient $C$, shortest path length ${\it l}$, and Pearson correlation coefficient $r$ for the visibility graph of the 
	magnitude time series with $N=2^{12}$. The synthetic catalog corresponds to the exponentially distributed 
	heights with $\lambda = 2.303$ (i.e., $b = 1$).}
\footnotesize	
\begin{tabular*}{\linewidth}{l@{\extracolsep{\fill}}ccccc}
\hline \hline \vspace{-0.27cm} \\
Region     &  $k_{\rm max}$ & $\langle k \rangle$ & $C$     & ${\it l}$  & $r$  \\ \vspace{-0.27cm} \\ \hline \hline
Tohoku1  &  101       & 6.76$\pm 0.07$      & 0.770$\pm 0.002$   & 5.49$\pm 0.03$ & 0.118$\pm 0.017$ \\ 
Tohoku2  &  82         & 6.36$\pm 0.17$      & 0.764$\pm 0.004$  & 5.66$\pm 0.04$ & 0.167$\pm 0.029$ \\
Kumamoto &  86        & 6.61$\pm 0.11$                & 0.769$\pm 0.004$    & 5.64$\pm 0.03$         & 0.128$\pm 0.016$   \\
Southern California   &  94        & 6.58$\pm 0.12$                  & 0.765$\pm 0.004$     & 5.64$\pm 0.02$       & 0.133$\pm 0.020$   \\ \hline 
Hakone                &  108       & 6.92$\pm 0.19$                  & 0.766$\pm 0.002$     & 5.28 $\pm 0.02$       & 0.118$\pm 0.008$   \\ 
Izu                        &  110        & 6.69$\pm 0.13$                  & 0.762$\pm 0.002$     & 5.80$\pm 0.02$        & 0.125$\pm 0.017$  \\ \hline
Cascadia              &  109       & 6.88$\pm 0.18$               & 0.751$\pm $0.002     & 5.84$\pm 0.03$         & 0.158$\pm 0.032$   \\
Shikoku                &  129       & 7.05$\pm 0.12$                  & 0.759$\pm 0.002$     & 5.43$\pm 0.02$         & 0.092$\pm 0.019$   \\  \hline
Synthetic Catalog     &  82        & 6.64$\pm 0.05$                  & 0.780$\pm 0.002$     & 5.67$\pm 0.02$         & 0.122$\pm 0.011$  \\ \hline \hline
\end{tabular*}
\label{tab:mag}
\end{table*}   
\begin{table*}
	\caption{Average values of the maximum degree $k_{\rm max}$, average degree $\langle k \rangle$, clustering
	coefficient $C$, shortest path length ${\it l}$, and Pearson correlation coefficient $r$ for the visibility
	graph of the inter-event time series with $N=2^{12}$. The data for
        swarms and tremors show disassortative degree mixing.  The last column represents
        the values of the degree distribution exponent $\gamma$ obtained from the entire IET series.}
\footnotesize        
\begin{tabular*}{\linewidth}{l@{\extracolsep{\fill}}cccccc}
\hline \hline \vspace{-0.27cm} \\
Region          &  $k_{\rm max}$ & $\langle k \rangle$ & $C$     & ${\it l}$  & $r$  & $\gamma$ \\ \vspace{-0.27cm} \\ \hline \hline
Tohoku1       &  435       & 8.52$\pm 1.09$               & 0.785$\pm 0.003$    & 4.99$\pm 0.05$        &-0.008$\pm 0.090$  & 2.34$\pm 0.05$ \\
Tohoku2       &  148       & 7.01$\pm 0.35$                 & 0.782$\pm 0.004$    & 5.54$\pm 0.03$        & 0.097$\pm 0.043$  & 2.60$\pm 0.08$\\ 
Kumamoto    &  477       & 8.71$\pm 3.35$                  & 0.780$\pm 0.013$     & 5.23$\pm 0.09$         & 0.021$\pm 0.169$   & 1.73$\pm 0.08$ \\
Southern California   &  188       & 7.20$\pm 0.49$                  & 0.784$\pm 0.003$     & 5.32 $\pm 0.03$        & 0.071$\pm 0.048$   & 2.64$\pm 0.05$ \\ \hline 
Hakone                &  1750      & 17.06$\pm 1.80$                 & 0.790$\pm 0.015$     & 3.24$\pm 0.04$        & -0.211$\pm 0.046$    & 1.81$\pm 0.09$ \\ 
Izu                        &  1714      &  15.99$\pm 2.74$                &  0.796$\pm 0.009$    & 3.55$\pm 0.03$         & -0.223$\pm 0.068$   & 1.79$\pm 0.09$ \\ \hline
Cascadia              &  701       & 11.89$\pm 1.30$                & 0.816$\pm 0.006$   & 3.98$\pm 0.04$         &-0.107$\pm 0.028$    & 2.51$\pm 0.05$ \\
Shikoku                &  1185       & 13.78$\pm 0.72$                & 0.828$\pm 0.006$     & 3.45$\pm 0.05$         &-0.162$\pm 0.021$    & 2.13 $\pm 0.05$ \\  \hline \hline
\end{tabular*}
\label{tab:IET}
\end{table*}      

\section{Discussion}
\label{sec:discussion}

   Finding and characterizing any correlations in the time series of earthquake magnitude is a subject of great importance
   as it may be useful in forecasting major earthquakes. However, to date, existence of correlations
   is somewhat controversial and has not been settled~\citep{Corral2006,Lippiello2008,Davidsen2011,Lippiello2012}.
   For instance, it was reported that regular earthquakes occurring close in space and time are correlated
   in their magnitudes~\citep{Lippiello2008}. A counterargument was given in Ref.\ \citep{Davidsen2011} that
   these were pseudo correlations due to the magnitude incompleteness and the modified Omori law. 
   To shed new light to this long-standing problem, we have made use of the complex network theory
   and analyzed the visibility graph to extract correlations in magnitude time series.
   
   The previous studies~\citep{Telesca_2012,Aguilar2013} in this context involve regular earthquakes 
   only. Here we extend the analysis to two other types of earthquakes~\citep{Obara2016} to consider this 
   problem in a more general perspective. By using the method of visibility graph, we have analyzed seismic 
   time series in seven seismogenic zones including the regular earthquakes in Southern California in common 
   with Ref.\ \citep{Aguilar2013} but for more extended time period. The degree distribution appears to be fitted 
   with a stretched exponential function for all the types of earthquakes analyzed here.
   
   Visibility graphs are also constructed from shuffled catalogs (Fig.\ \ref{fig:degdistSH}) or 
   synthetic data drawn randomly from the GR law (Fig.\ \ref{fig:degdistEXP}).
   On average, the degree distribution appears to be fitted with the stretched exponential function.
   However, the Kolmogorov-Smirnov test rejects the null hypothesis that these degree distributions are identical.
   Namely, the degree distributions for these surrogate data are indeed distinguishable from that of the original data.
   This means that the original series have some special characters that are lost in their surrogates: shuffled or synthetic catalogs.

   Since the criterion for the visibility graph involves both magnitude and IET, one might argue that 
   the difference in the degree distribution detected by the KS test is a mere by-product resulting from the correlation in IET.
   To exclude this possibility, we also perform the KS test by constructing the visibility graph using the event index $i$
   instead of the occurrence time $t_i$.
   We find that the null hypothesis is again rejected.
   Namely, the magnitude series $(i, M_i)$ leads to slightly different degree distributions if they are shuffled,
   although the difference is detectable only by the KS test.
   Thus, the memory should exist in magnitudes alone.
   
   The memoryless nature of earthquake magnitudes is a basic assumption in the epidemic-type aftershock 
   sequences (ETAS) model, which is the most successful statistical model for earthquake time series 
   \citep{Ogata1988}. The results given here implies that the memoryless assumption in earthquake magnitude
   is rather approximate. Thus, if one wishes to improve statistical models for earthquake occurrence, 
   the correlation in magnitude should be taken more seriously. To this end, the correlation found here
   should be defined and quantified more clearly.

   The degree distributions of stretched exponential form appear to contradict some previous studies 
   \citep{Telesca_2012,Aguilar2013}, in which the power law tails are concluded for the magnitudes of 
   regular earthquakes. In view of Eq.\ (\ref{eq:hurstexpo}), this may imply a fBm-like correlation in 
   the magnitude time series. Interestingly, however, they also analyzed randomly shuffled sequences 
   of magnitudes and did not find any significant difference in the degree distributions. This rather 
   contradicts the existence of a fBm-like correlation. Additionally, the degree distribution obtained in 
   Ref.\ \citep{Telesca_2012} spans approximately one decade only, and the tails are noisy. Thus, one 
   needs to be careful to draw a conclusion based on these data alone. In Ref.\ \citep{Aguilar2013}, the 
   tails of the degree distributions are less noisy, but they appear to fall off from a power law at their 
   tails. Thus, their degree distributions might be fitted with a stretched exponential function. However, 
   the degree distribution produced from Mexican catalog appears to develop a tail that is still different 
   from stretched exponential. We noticed that the magnitude data in the Mexican catalog do not always 
   obey the GR law, and this may be the reason for the deviation from the stretched exponential function.
   However, the Mexican data require more careful and dedicated analyses to draw any decisive 
   conclusions on specific type of magnitude correlation.
   
   We apply the visibility-graph analysis for the characterization of the inter-event times (IET)
   between consecutive earthquakes. Contrary to the magnitude time series, 
   we find an evidence of fBm-like correlations between the inter-event times. The network associated with 
   the IET series has a scale-free nature with the exponents $\gamma$, which depends on the essential 
   characteristics of seismic activity. In the context of the $f^{-\beta}$ noise, the exponent $\gamma$ is 
   directly related to $\beta$. These exponents may work as a generalized and unified quantification of the 
   intermittent nature of seismic time series. For instance, we find that the IET series for swarms are 
   similar to superdiffusive Brownian motion, whereas those for regular earthquakes correspond subdiffusion. 
   However, the interpretation of superdiffusive or subdiffusive nature in the IET series is yet unclear from the 
   mechanical point of view, and should be pursued in the subsequent studies.
       
   We have also analyzed the whole set of data using the horizontal visibility graph 
   algorithm. For both magnitude and IET series, however, the degree distribution results in an 
   exponential distribution and no appreciable change has been observed between these different data 
   sets, making it harder to draw any conclusive remarks on the distinction of different time series.

\section{Conclusion}
\label{sec:summary}

   In conclusion, we have investigated the correlations in the time series of magnitudes and of inter-event times (IETs)
   for three different categories of earthquakes in seven seismogenic zones in the world.
   By applying the methods of visibility graph, we show that the IET series possess correlations similar to fractional Brownian motion,
   and that the three categories of earthquakes have different exponents.
   While such an apparent correlation is absent in the magnitude series, the Kolmogorov-Smirnov test on the degree distribution
   reveals that the earthquake magnitudes are not statistically equivalent to an uncorrelated (random or shuffled) time series.
   This challenges a current popular belief that magnitude time series are random.
   Since current major statistical models for earthquake rate are based on this belief,
   these results provide us with useful constraints in developing better statistical models.
   
   Different temporal behaviors of three categories of earthquakes are also reflected in various graph-theoretical quantities.
   As found from the analysis of the assortativity coefficient, the swarms are more intermittent than tremors.
   More graph-theoretical techniques including horizontal visibility graph~\citep{Luque2009}, multiplex visibility graph 
   \citep{Lacasa2015}, and recurrence networks~\citep{Donner_2010}
   would give new criteria for categorizing or unifying different seismic activities. A novel approach for forecasting
   time series based on visibility graph~\citep{Zhao2020} might find potential application for earthquakes.
   Our study therefore shows with affirmation that the visibility graph algorithm has the potentiality 
   to capture the non-trivial complexity inherent in a time series which is nonlinear and nonstationary in nature.   

   One can also consider more elaborated methods for the graph construction~\citep{XU2018}. For instance, the visibility graph 
   constructed here is undirected and unweighted. Time directionality and weighted links based on the inter-event distances
   would be interesting subjects. Additionally, since the spatial information of the seismic events has 
   been disregarded here, the extension of the visibility graph method to space-time may be a promising attempt.
   
   Together with the present results, such graph-theoretical approaches would bring benefits to statistical modeling of various types of
   seismic activities that cannot be reproduced by the well-established ETAS model for regular earthquakes.


\section*{Conflict of Interest Statement}

The authors declare that the research was conducted in the absence of any commercial or financial relationships that could be construed as a potential conflict of interest.

\section*{Author Contributions}

SK and TH conceived and designed the study and drafted the manuscript. SK performed computer simulations and carried out 
the data analysis. AO and YY prepared the earthquake catalogs. All authors took part in discussing the results, reading and 
approving the final version of the paper.

\section*{Funding}

This study was supported by Japan Society for the Promotion of Science (JSPS) Grants-in-Aid for Scientific Research (KAKENHI) Grants
Nos. JP16H06478 and 19H01811. Additional support from the MEXT under 
“Exploratory Challenge on Post-K computer” (Frontiers of Basic Science: Challenging the Limits) and the 
“Earthquake and Volcano Hazards Observation and Research Program” is also gratefully acknowledged.


%

\end{document}